\documentclass{article}

 \usepackage[final]{neurips_2019}
 \usepackage{authblk}

\usepackage[utf8]{inputenc} 
\usepackage[T1]{fontenc}    
\usepackage{hyperref}       
\usepackage{url}            
\usepackage{booktabs}       
\usepackage{amsfonts}       
\usepackage{nicefrac}       
\usepackage{microtype}      
\usepackage{graphicx}
\usepackage{subcaption}     
\title{Hepatocellular Carcinoma Intra-arterial Treatment Response Prediction for Improved Therapeutic Decision-Making}

\author[1]{ Junlin Yang}
\author[1]{ Nicha C. Dvornek}
\author[1]{ Fan Zhang}
\author[1]{Julius Chapiro}
\author[1]{MingDe Lin}
\author[2]{Aaron Abajian}
\author[1]{James S. Duncan}
\affil[1]{Yale University, New Haven, CT, USA}
\affil[2]{University of Washington, Seattle, WA, USA }

\begin{document}

\maketitle

\begin{abstract}

This work proposes a pipeline to predict treatment response to intra-arterial therapy of patients with Hepatocellular Carcinoma (HCC) for improved therapeutic decision-making. Our graph neural network model seamlessly combines heterogeneous inputs of baseline MR scans, pre-treatment clinical information, and planned treatment characteristics and has been validated on patients with HCC treated by transarterial chemoembolization (TACE). It achieves Accuracy of $0.713 \pm 0.075$, F1 of $0.702 \pm 0.082$ and AUC of $0.710 \pm 0.108$. In addition, the pipeline incorporates uncertainty estimation to select hard cases and most align with the misclassified cases. The proposed pipeline arrives at more informed intra-arterial therapeutic decisions for patients with HCC via improving model accuracy and incorporating uncertainty estimation.

\end{abstract}

\section{Introduction}
Hepatocellular carcinoma (HCC), primary liver cancer, has the fastest rising incidence rates worldwide, especially in the western countries \cite{siegel2019cancer}. Transarterial chemoembolization (TACE) has been a well established primary therapy for patients with unresectable HCC \cite{lencioni2013chemoembolization}. The assessment of patients after TACE treatment has been advanced by the quantitative European Association for the Study of the Liver (qEASL) response criterion, which quantitatively measures the degree of change in 3D enhancing tumor volume instead of 2D measurement or visual estimation. However, it still remains a clinical challenge to predict which patients will respond to TACE before treatment \cite{mannelli2013serial}.

Why does early prediction before treatment matter? Data-driven methods for medical imaging have been emphasized more on the analytical models such as segmentation and classification for diagnostic purposes compared to predictive models \cite{zhang2018liver}. However, the advancement in predictive models to predict future medical observations, such as disease progression \cite{akbari2016imaging}, survival and prognosis \cite{macyszyn2015imaging}, and treatment response \cite{mannelli2013serial}, with high precision could impact the development of treatment procedures and could modify treatment strategy. 
Here, we focus on treatment response prediction of the TACE procedure. Predicting and identifying non-responders to TACE prior to initiation of therapy carries significant potential survival benefits for non-responders, should such patients be allowed to enter alternative, e.g. systemic, therapies. 

Previous work on treatment response relies heavily on clinical features and handcrafted radiomics features \cite{mannelli2013serial}\cite{abajian2018predicting}. Recently, more and more convolutional neural networks that take images as inputs have been utilized for treatment response prediction \cite{shi2019machine}\cite{wu2019deep}\cite{Peng2019}. However, integration of readily-available non-imaging data, such as clinical information and treatment characteristics, would likely improve prediction accuracy. This paper leverages recent advances in graph neural networks, which has shown the power of handling heterogeneous inputs. In medical imaging, it has been widely used for brain data analysis, e.g., to handle imaging and non-imaging data \cite{parisot2017spectral} and to identify biomarkers from complicated relationships \cite{li2019graph}.

This paper presents the first work to explore graph neural networks for prediction of treatment response. Some level of uncertainty exists not only in the model and data, but also in the ground truth labels (as illustrated in Sec. 2). The pipeline incorporates uncertainty estimation to select difficult cases that are often misclassified.

\section{Method}
\textbf{Problem formulation} Pre-treatment baseline data from HCC patients are collected. We aim to predict TACE treatment response (one month follow-up) from baseline data. Since treatment response can be assessed by changes in qEASL value \cite{mannelli2013serial} and over a 65\% reduction in qEASL between baseline and follow-up imaging indicates responders, the prediction problem can be conceptualized as a classification problem. 

\begin{figure}[!t]
    \centering
    \begin{minipage}{.54\textwidth}
        \centering
        \includegraphics[width=\linewidth]{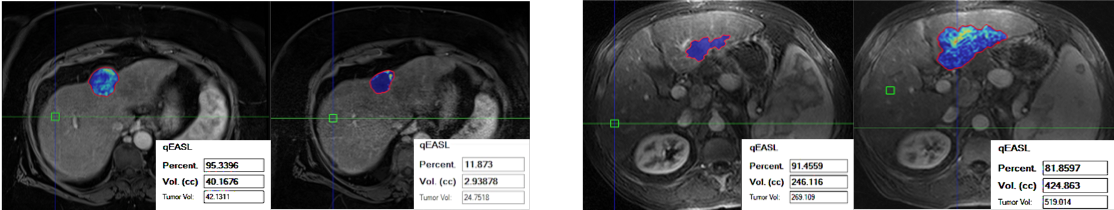}
        \caption{Two examples of qEASL analysis, left is from responder, right is from non-responder. For each example, left is baseline scan, right is follow-up scan. We can see the qEASL value drops from 40.17 to 2.94 cm$^3$ for responder, while the qEASL value drops from 246.12 to 424.86 cm$^3$ for non-responder. Similar estimation is performed three times for each patient and averaged to generate the final ground truth label.}
        \label{fig:fig1}
    \end{minipage}%
      \hspace{0.02\textwidth}
    \begin{minipage}{0.42\textwidth}
        \centering
        \includegraphics[width=\linewidth]{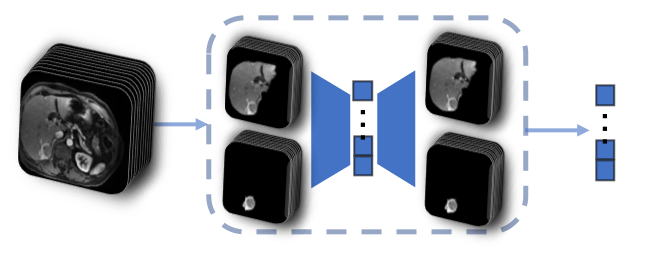}
        \caption{Generating node feature vectors from 3D volumes using autoencoder.}
        \label{fig:fig2}
    \end{minipage}
\end{figure}

\textbf{Data} The dataset consists of 83 patients with HCC treated by TACE. Both baseline and one month follow-up multi-phasic MR scans are collected. Non-imaging data includes pre-treatment clinical information (laboratory values, clinical history, etc) and planned treatment characteristics. As mentioned above, 65\% drop in qEASL indicates responders. To generate ground truth labels, qEASL analysis was performed on both baseline MR and one month follow-up MR scans and changes were computed accordingly. 20-second arterial phase images were selected for qEASL analysis, as shown in Fig. \ref{fig:fig1}. qEASL values are essentially the enhancing tumor volume expressed as a percentage of the total tumor volume. To estimate qEASL, each measurement includes three parenchymal regions of interest (ROIs) to generate an average, serving as the estimated parenchymal intensity.


\begin{figure}
\begin{center}
\includegraphics[width=0.9\linewidth]{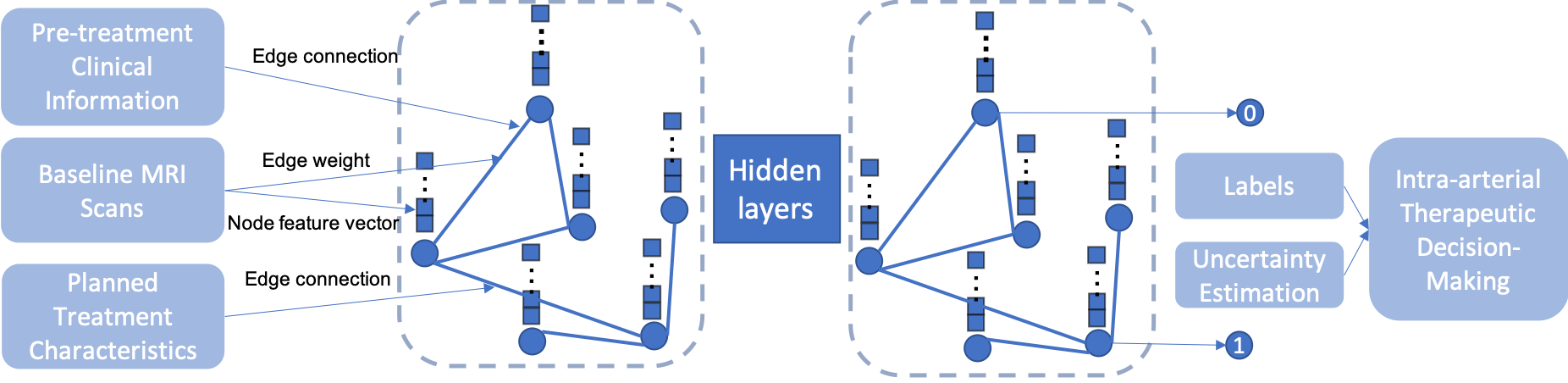}
\end{center}
   \caption{Pipeline for predicting treatment response using GCN and uncertainty estimation.}
\label{fig:pipeline}
\end{figure}

\textbf{Pipeline}
The proposed prediction pipeline consists of three steps, as shown in Fig. \ref{fig:pipeline}. First, build the graph using both imaging and non-imaging data, where each patient serves as a node. Second, train the graph convolutional neural network (GCN) with softmax for semi-supervised classification on the above graph to get prediction results. Third, use Monte Carlo (MC) dropout as Bayesian estimation for uncertainty estimation to identify hard cases for more informed decision-making. 

To build the graph, node feature vectors encoding imaging information are generated by a 3D autoencoder (AE) model as shown in Fig. \ref{fig:fig2}. The AE model is fed concatenated liver and tumor 3D volumes and trained on the self-reconstruction task. The AE latent vectors of length of 128 are extracted as node feature vectors. Graph edges incorporate the non-imaging data. According to the prior knowledge from physicians, two binary features from both clinical information (Cirrhosis presence) and treatment characteristics (Sorafenib) are selected. An edge is drawn for each binary feature whose status is shared between two patients to form the adjacency matrix. Correlations between each pair of nodes are computed to be applied as weights on the above adjacency matrix.

The graph convolutional neural network (GCN) \cite{kipf2016semi} was trained on the graph built above. The structure of the GCN consists of convolutional layers, Rectified Linear Units (ReLU), and a softmax activation function at the end. To avoid over-fitting and realize uncertainty estimation, a dropout rate of 0.15\% was applied during both training and testing. To train the prediction model, the cross-entropy loss function was calculated only over labelled training nodes during training stage, and then used for updating the parameters in the GCN. During testing stage, unlabelled testing nodes are assigned labels according to the output of the softmax.

To generate the final prediction and corresponding uncertainty estimation, MC dropout \cite{gal2016dropout} was utilized. For each sample, predictions were performed 100 times using the GCN model with dropout. The final prediction was decided by majority voting.  Confidence of the prediction was estimated quantitatively by the ratio of predictions that agreed with the final prediction. The confidence level for each prediction can be used to select the most uncertain cases, as they are likely to be the difficult cases that are more likely to be misclassified.

\section{Results and Analysis}

\textbf{Classification results} 10-fold cross-validation was applied for evaluation. Please refer to Table \ref{sample-table} for details. GCN is the graph convolutional neural network model in the above proposed pipeline. RF is a random forest model with the same imaging and binary non-imaging features as inputs where PCA was used for dimensional reduction. Ablation 1-3 w/o Cirrhosis/Sorafenib/non-imaging refers to building the graph without Cirrhosis/Sorafenib/both non-imaging features. GCN shows a significant improvement in prediction performance compared to the random forest model. The ablation studies show that the proper construction of the graph with prior knowledge is essential for the success of GCN. The drop in performance for each ablation study corresponds to the importance of the dropped non-imaging feature. 

\begin{table}
  \caption{Comparison of prediction performance}
  \label{sample-table}
  \centering
  \scalebox{0.75}{
  \begin{tabular}{l|l|l|l|l|l}
    \toprule
    Method   &Accuracy (std)  & F1 (std)   &   AUC (std) \\
    \midrule
    RF & $0.58 \pm 0.06$ &$ 0.57    \pm 0.08 $& $0.60 \pm 0.08$  \\
    GCN    &   $\textbf{0.71} \pm \textbf{0.07}$     &$\textbf{0.70}\pm\textbf{0.08}$& $\textbf{0.71} \pm \textbf{0.10}$   \\
    Ablation 1 w/o Cirrhosis &$0.64\pm0.05$& $0.62\pm0.06$ &$0.66\pm0.09$\\
    Ablation 2 w/o Sorafenib &$0.65\pm0.05$&$0.63 \pm 0.07$&$0.59\pm 0.07$\\
    Ablation 3 w/o non-imaging &$0.59 \pm 0.03$&$0.57 \pm 0.06$&$0.63\pm 0.11$ \\
    \bottomrule
  \end{tabular}}
\end{table}

\textbf{Uncertainty estimation} Uncertainty estimation \cite{gal2016dropout} was achieved by MC dropout during the test stage. By ruling out test cases with the lowest confidence, we can see the classification performance generally improves, which shows that the majority of low confidence cases align with misclassified cases. Specifically, when ruling out cases with confidence lower than 85\%, 90\%, 95\%, computed on remaining cases, F1 improves by 3.76\%, 3.6\%, 5.76\%, AUC improves by 2.11\%, 3.56\%, 8.61\%, and Accuracy improves by 5.45\%, 6.58\%, 10.61\%.

\section{Conclusion}

In summary, the proposed pipeline arrives at more informed intra-arterial therapeutic decisions for HCC patients via improving model accuracy and incorporating uncertainty estimation. GCN incorporates prior knowledge into the graph construction and combines both imaging and non-imaging features. Uncertainty estimation serves as an essential role towards more informed clinical decision-making. Yet, much remains to be improved. For future research, more flexible graph construction such as constructing multiple graphs instead of one graph could help incorporate more prior information. Other uncertainty estimation methods should also be investigated. 

\small
\bibliographystyle{unsrt}
\bibliography{reference}

\end{document}